\documentclass[aps,twocolumn,prc,amssymb,amsmath,bm,floatfix,superscriptaddress,nofootinbib,float]{revtex4}
\usepackage[dvipsnames,usenames]{xcolor}
\usepackage{orcidlink}
\usepackage{graphicx}
\usepackage{mathptmx}
\usepackage{bm}

\newcommand{\be}{\begin{equation}}
\newcommand{\ee}{\end{equation}}
\newcommand{\bea}{\begin{eqnarray}}
\newcommand{\eea}{\end{eqnarray}}
\newcommand{\ba}{\begin{align}}
\newcommand{\ea}{\end{align}}

\newcommand{\rme}{{\rm e}}
\newcommand{\rmi}{{\rm i}}

\newcommand{\rs}{{\rm s}}
\newcommand{\rn}{{\rm n}}
\newcommand{\rp}{{\rm p}}

\newcommand{\para}{\parallel}

\begin{document}

\title{Superfluid density in disordered pasta phases in neutron star crusts}
\author{Zhao-Wen Zhang\orcidlink{0000-0002-3249-344X}}
\affiliation{The Niels Bohr International Academy, The Niels Bohr Institute, University of Copenhagen, Blegdamsvej 17, DK-2100 Copenhagen \O, Denmark}
\affiliation{NORDITA, KTH Royal Institute of Technology and Stockholm University, Hannes Alfv\'ens v\"ag 12, SE-106 91 Stockholm, Sweden}
\author{C. J. Pethick  \orcidlink{0000-0003-0106-7891}  }
\affiliation{The Niels Bohr International Academy, The Niels Bohr Institute, University of Copenhagen, Blegdamsvej 17, DK-2100 Copenhagen \O, Denmark}
\affiliation{NORDITA, KTH Royal Institute of Technology and Stockholm University, Hannes Alfv\'ens v\"ag 12, SE-106 91 Stockholm, Sweden}
\begin{abstract}
We  calculate the superfluid density of nucleons in disordered pasta phases in the inner crust of neutron stars using an effective medium approach which parallels that previously used for calculating the electrical conductivity of terrestrial matter. We allow for the effect of entrainment, the fact that the current density of one species of nucleon depends on the gradient of the phase not only of the same species but also of the other species.    The superfluid density tensors for the perfectly ordered pasta phases are very anisotropic, and we derive expression for the effective superfluid densities of the disordered phases.
\end{abstract}
\maketitle

\section{Introduction}
The superfluid density of neutrons in the crust of the star is an important property in models of glitches in neutron star rotation rates, and the corresponding quantity for protons is an important ingredient  in calculations of magnetic properties.
At densities approaching nuclear saturation density it is predicted that phases with nonspherical nuclei can occur in the crust of neutron stars \cite{Lorenz, OyamatsuNS}.  Whether or not such phases have lower energy than  a uniform mixture of neutrons, protons and electrons is still unclear, since the result depends on the nuclear interaction employed \cite{Lorenz,DouchinHaensel}.   Further work with improved calculational methods and nuclear interactions is needed to determine whether these phases are thermodynamically stable.

In the crust at densities above that for neutron drip, neutrons and protons are predicted to be superfluid.  In the phase with round nuclei, flow of neutrons can occur over large distances because of the neutrons outside nuclei.  The protons, however, cannot participate in bulk flows because individual nuclei are well separated in space, and there is negligible proton tunneling between nuclei.  The phases with spherical nuclei are predicted to have cubic symmetry, and consequently the neutron superfluid density tensor is isotropic. For the pasta phases, which have nonspherical nuclei, superfluid flow of protons is possible because spaghetti strands and lasagna sheets are extended.  For perfect pasta phases with long-range spatial order, the superfluid density tensors for both neutrons and protons are anisotropic, with different values for directions parallel to and perpendicular to the symmetry axis of the pasta, the direction of the spaghetti strands and the normal to the lasagna sheets.

We turn now to disordered phases.  On length scales large compared with the lattice spacing and the length scale on which the orientation of the crystal axes changes, the effective superfluid density tensor is isotropic if the orientation of the axes is random.  The phase with spherical nuclei has
cubic symmetry. For this case, the superfluid density tensor is isotropic and,  to the extent that effects due to the boundary between regions with different crystal orientations are negligible, the effective superfluid density tensor is equal to that in the perfectly ordered medium.

In this paper we calculate the effective superfluid density for disordered lasagna and spaghetti phases.  We adopt an effective medium approach inspired by earlier work on the effective electrical conductivity  \cite{Bruggeman,Landauer, Stroud, BergmanStroud}.  Such methods have also been applied to calculate elastic properties  of polycrystals \cite{Berryman}, including those in neutron star crusts \cite{KobyakovCJP, CJPZhangKobyakov}. The basic idea in these methods may be described as follows.  One considers an inclusion with anisotropic properties embedded in a homogeneous, isotropic medium and then demands that, on averaging over possible orientations of the inclusion, the response of the system is the same as that of the medium outside the inclusion.  Another way of expressing this in terms of scattering is that the properties of the medium are chosen so that, on average, the inclusion gives no scattering of an incident disturbance.

This article is organized as follows.  Section II sets out the basic formalism, and a number of details are described in the Appendix.  Applications are given in Sec.\ III, which begins with the case of a single component and then goes onto the two-component case with entrainment.   In Sec. IV we discuss applications of our results to the spaghetti and lasagna phases and the effects of magnetic fields on the protons.  Concluding remarks are made in Sec.\ V.

\section{Basic formalism}

Consider a perfectly ordered system with a spatially periodic structure and made up of neutrons and protons (which are both superfluid), and normal electrons.  In a generalization of the two-fluid model for liquid helium II  we may write the current density of nucleons for long-wavelength, low-frequency phenomena in the form
\be
j_\alpha^i= \frac{n_{\alpha\beta}^{\rs, ij}}{m} \left(\nabla^j\phi_\beta-\frac{q_\alpha A^j}{c}\right) +n_\alpha^{\rn,ij} u^j.
\label{currentalpha}
\ee
Here the Greek subscripts $\alpha$ and $\beta$ refer to the nucleon species (n for neutrons and p for protons), the indices $i$ and $j$ refer to Cartesian coordinates and  $\phi_\alpha$ is equal to one half of the phase of the condensate pair wave function for species $\alpha$ averaged over distances large compared with the lattice spacing and other microscopic length scales but small compared with the characteristic length scale of the phenomenon in question.     The components of the normal fluid velocity, which is the velocity of the periodic structure of the medium, are denoted by $u^j$, and those of the vector potential by $A^j$, and $q_\alpha$ is the charge of species $\alpha$.  The quantity $n_{\alpha\beta}^{\rs, ij}$ is the superfluid density tensor, and it is in general not diagonal in the species variables because of entrainment of the two superfluids, the fact that a gradient in the phase of one component can give rise to a flow of the other component. The normal density tensor is denoted by $n_\alpha^{\rn,ij}$.  We shall work in units in which $\hbar=1$.  We shall assume that the electrons are at rest in the frame moving with the normal velocity, but they will not enter in our subsequent considerations.

In most of the paper we shall neglect the effects of magnetic fields, so we shall take the vector potential to be zero.  For small $\nabla^j\phi_\beta$ and small $u^j$, the dependence of the superfluid and normal density tensors on these variables may be neglected.  However, as we shall describe in Sec.\ IV, we shall comment on the effect of magnetic fields on the protons.

We shall confine our attention to systems in which the principle axes of the medium are, or may be chosen to be, orthogonal.  This applies to all crystal systems apart from the triclinic and monoclinic ones, and is thus general enough to cover the cases of interest in neutron star crusts.  For the lasagna phase with uniform sheets, the principle axes may be taken to be the normal to the sheets and two orthogonal axes lying in the plane of the sheets.  For the spaghetti phase, one expects the rods to be arranged in a hexagonal pattern, and thus the superfluid and normal density tensors are isotropic in the plane perpendicular to the strands.  Sheets in the lasagna phase may be spatially modulated in the plane of the sheets but,  according to quantum molecular dynamics calculations \cite{SchneiderBerryBriggsetalWaffles},  the modulation has hexagonal symmetry, and therefore also this case falls into this class.       It is simplest to work in terms of components of vectors along the principle axes of the medium, in which case we may write Eq.\ (\ref{currentalpha}) in the form
\be
j_\alpha^\lambda= \frac{n_{\alpha\beta}^{\rs \lambda}}{m} \nabla^\lambda\phi_\beta +n_\alpha^{\rn\lambda} u^\lambda,
\label{currentalpha}
\ee
where $\lambda$ denotes the principle axes and there is no sum over $\lambda$ on the right side of the equation.

In a system consisting of randomly oriented domains, one expects the spatial average of the current density to be proportional to the spatial average of the gradient of the phases and to be in the same direction, and thus, omitting the magnetic field term, one may write
\be
\langle j_\alpha^i\rangle= \frac{n_{\alpha\beta}^{\rs, \rme}}{m} \langle\nabla^i\phi_\beta\rangle +n_\alpha^{\rn,\rme} u^i,
\label{currentalphaav}
\ee
where the angular brackets denote a spatial average, which we shall assume to be equivalent to an ensemble average.  The objective of this article is to calculate the spatially isotropic quantity $n_{\alpha\beta}^{\rs, \rme} $, the effective superfluid density tensor for the disordered medium.   The effective normal density is denoted by $n_\alpha^{\rn,\rme}$.  Under a Galilean transformation to a reference frame moving at a velocity $-\vec v$ with respect the original frame, $\vec \nabla \phi_\beta$ changes by $m\vec v$ and the total current density of species $\alpha$ changes by $n_\alpha \vec v$, where $n_\alpha$ is the density of the species.  It therefore follows from Eq.\ (\ref{currentalphaav}) that
\be
\sum_\beta n^{\rs,\rme}_{\alpha\beta} +n^{\rn, \rme}_\alpha =n_\alpha.
\label{Galileo}
\ee

As we demonstrate in the Appendix (see Eqs.\ (\ref{f}) and (\ref{selfconsist})), the effective superfluid density in the effective medium approach is given by solving the matrix equation
\begin{align}
&\sum_{\lambda=1,2,3} 3 n^{\rs,\rme}  (2n^{\rs, \rme}+ n^{\rs \lambda})^{-1} (n^{\rs \lambda} - n^{\rs, \rme}  )  \nonumber \\
=&\sum_{\lambda=1,2,3}  [{\cal I}+ (n^{\rs \lambda} - n^{\rs, \rme}  ) (3n^{\rs,\rme})^{-1}]^{-1}  (n^{\rs \lambda} - n^{\rs, \rme}  )=0,
\label{final2}
\end{align}
where the symbols $n$ with no subscripts denote $2\times2$ matrices in the space of nucleon species (neutrons and protons) and $\cal I$ is the unit matrix.  The form (\ref{final2}) exhibits explicitly the fact that the quantity is symmetric in the species variables. It also has the form to be expected from a multiple scattering or screening problem, with the bare scattering ``potential'' being proportional to $n^{\rs \lambda} - n^{\rs, \rme}$ \cite{Stroud}.

\section{Applications}
We now apply our result to uniaxial systems, which have  one principle axis along the axis of the system for which quantities are denoted by $\para$, and two principle axes perpendicular to the  axis of the system, for which quantities are denoted by $\perp$. Equation (\ref{final2}) then reduces to
\begin{align}
n^{\rs,\rme}  (2n^{\rs, \rme}+ n^{\rs \para})^{-1}  (n^{\rs,\rme} -n^{\rs \para})\hspace{5em}\nonumber \\ + 2n^{\rs,\rme}  (2n^{\rs, \rme}+ n^{\rs \perp})^{-1}  (n^{\rs,\rme} -n^{\rs \perp})=0.
\label{uniaxial}
\end{align}

\subsection{No entrainment}

For uniform matter, the entrainment parameter $n_{\rn\rp}^{\rs}$ is not known very well, but estimates indicate that it is less than  $0.1 n_\rp$ \cite{KPRS}, and we expect that the entrainment parameters in the pasta phases will be correspondingly small.  Consequently, in calculating the pp and nn components of the effective superfluid density tensor, it is a good first approximation to neglect entrainment.  In that case,
 Eq.\ (\ref{uniaxial}) reduces to the two equations
\be
\frac{n_{\rn\rn}^{\rs,\rme} -n_{\rn\rn}^{\rs \para}} {2n_{\rn\rn}^{\rs, \rme}+ n_{\rn\rn}^{\rs \para}}  + 2 \frac{n_{\rn\rn}^{\rs,\rme} -n_{\rn\rn}^{\rs \perp}} {2n_{\rn\rn}^{\rs, \rme}+ n_{\rn\rn}^{\rs \perp}}=0,
\label{nen}
\ee
and
\be
\frac{n_{\rp\rp}^{\rs,\rme} -n_{\rp\rp}^{\rs \para}} {2n_{\rp\rp}^{\rs, \rme}+ n_{\rp\rp}^{\rs \para}}  + 2 \frac{n_{\rp\rp}^{\rs,\rme} -n_{\rp\rp}^{\rs \perp}} {2n_{\rp\rp}^{\rs, \rme}+ n_{\rp\rp}^{\rs \perp}}=0.
\label{nep}
\ee
These equations have the same form as that for the electrical conductivity of polycrystals and binary metallic mixtures \cite{Bruggeman, Landauer,Stroud}, the essential reason being that the electrostatic potential and the phase of the condensate wave function both satisfy the Laplace equation, except at the surface of the inclusion. Equations (\ref{nen}) and (\ref{nep}) may be written as
\be
2 {n_{\alpha\alpha}^{\rs,\rme}}^2-n_{\alpha\alpha}^{\rs,\rme} n_{\alpha\alpha}^{\rs\perp} -n_{\alpha\alpha}^{\rs\para} n_{\alpha\alpha}^{\rs\perp}=0,
\ee
for which the physically meaningful root is the positive one,
\be
n_{\alpha\alpha}^{\rs,\rme}=\frac{n_{\alpha\alpha}^{\rs\perp}+\sqrt{{n_{\alpha\alpha}^{\rs\perp}}^2+8n_{\alpha\alpha}^{\rs\para} n_{\alpha\alpha}^{\rs\perp}}}{4}.
\label{1component}
\ee

It is interesting to compare this result with those of simpler approaches to the problem that correspond to the Voigt and Reuss approximations in the theory of elastic properties of polycrystalline materials \cite{Voigt, Reuss}, which Hill demonstrated to give upper and lower bounds on the elastic constants \cite{Hill}.   In the Voigt approach one assumes that the strain is constant in all domains of the medium. For the superfluid density the corresponding assumption is that the gradient of the phase is the same in all domains, and  the effective superfluid density is the {\it arithmetic} mean  of the superfluid density tensor over possible orientations of the domains:
\be
(n^{\rs.\rme}_{\alpha\alpha})_{\rm Voigt}=\frac{ n^{\rs\para}_{\alpha\alpha}+2n^{\rs\perp}_{\alpha\alpha}       }{3}.
\ee
In the Reuss approach, the stress is assumed to be constant in all domains, and the analogous assumption for the superfluid density is that the superfluid current density is the same in all domains.  The effective superfluid density is the {\it harmonic} mean  of the superfluid density tensor over possible orientations of the domains:
\be
(n^{\rs.\rme}_{\alpha\alpha})_{\rm Reuss}=\frac{3}{ 1/n^{\rs\para}_{\alpha\alpha}+2/n^{\rs\perp}_{\alpha\alpha}       }  =\frac{3n^{\rs\para}_{\alpha\alpha}n^{\rs\perp}_{\alpha\alpha}}{ 2n^{\rs\para}_{\alpha\alpha}+n^{\rs\perp}_{\alpha\alpha}       }   .
\ee

\begin{figure}
\includegraphics[width=3.5in]{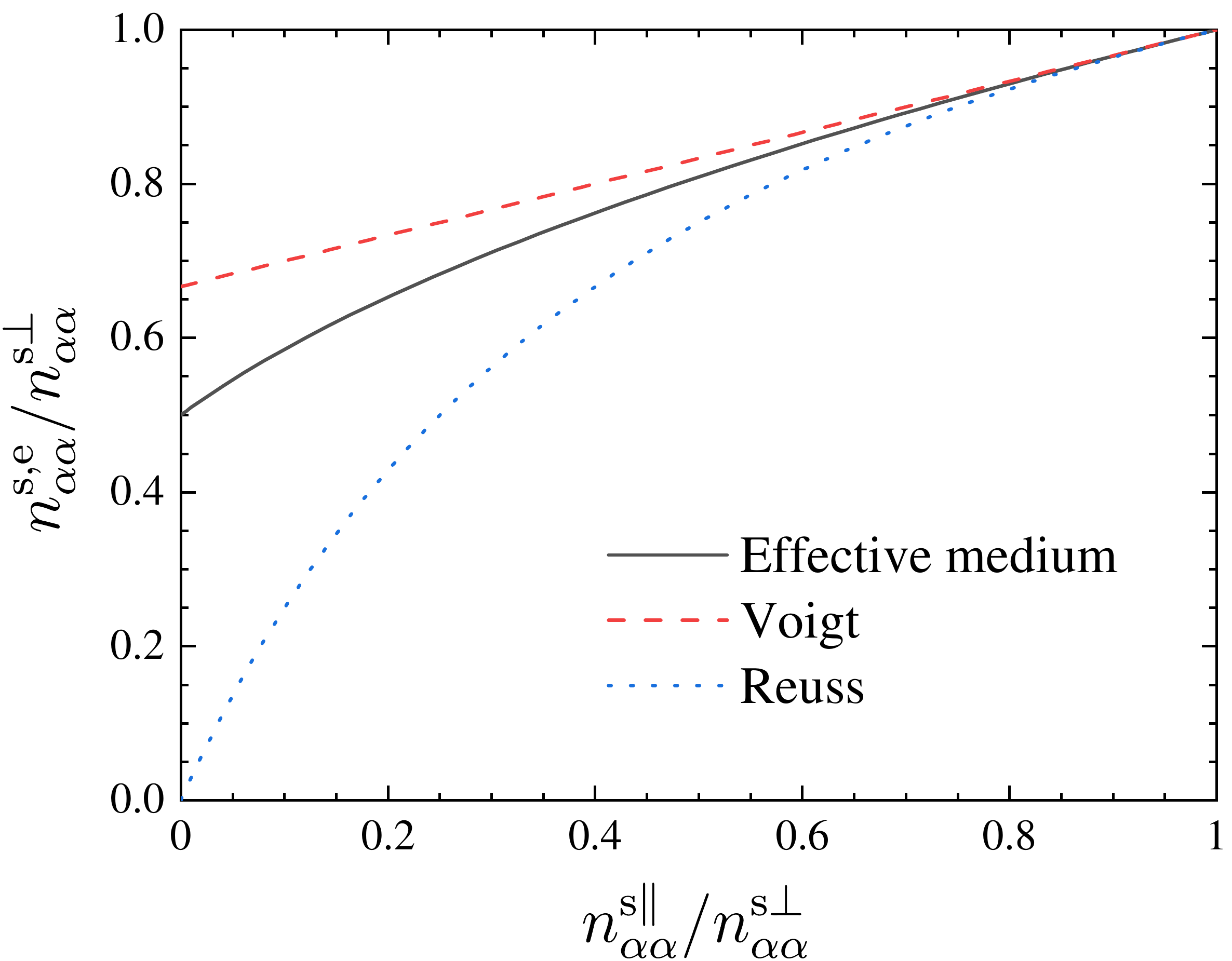}
\caption{Effective superfluid density component  $n^{\rs,\rme}_{\alpha\alpha}$ in units of $n^{\rs\perp}_{\alpha\alpha}$ as a function of $n^{\rs\para}_{\alpha\alpha}/n^{\rs\perp}_{\alpha\alpha}$ for the lasagna phase in the absence of entrainment.  The full line is the result of the effective medium theory, the dashed line is the Voigt approximation, and the dotted line is the Reuss approximation.}
\label{Lasagna}
\end{figure}

For the lasanga phase, one expects physically that superfluid flow is impeded in the parallel direction, and therefore $n_{\alpha\alpha}^{\rs\perp} >n_{\alpha\alpha}^{\rs\para}$ and we may write Eq.\ (\ref{1component})
as
\be
\frac{n_{\alpha\alpha}^{\rs\rme}}{n_{\alpha\alpha}^{\rs\perp}}=\frac{1+\sqrt{1+8n_{\alpha\alpha}^{\rs\para}/ n_{\alpha\alpha}^{\rs\perp}}}{4},
\ee
which is plotted in Fig.\ \ref{Lasagna}.   Remarkably, even if there is no superfluid flow perpendicular to the lasagna sheets, the predicted effective superfluid density falls to only one half of the value it would have if the parallel and perpendicular superfluid densities were equal.

\begin{figure}
\includegraphics[width=3.5in]{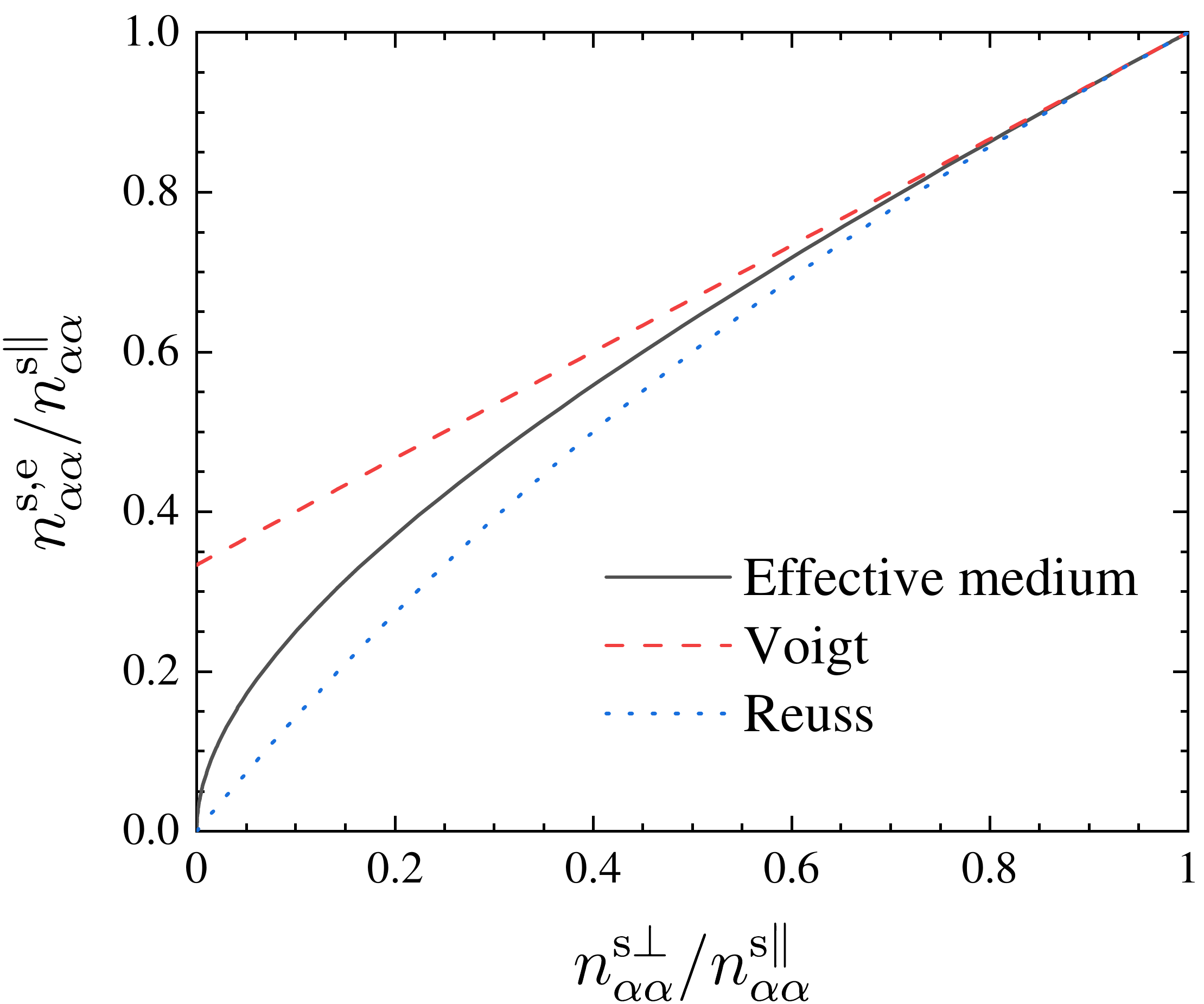}
\caption{Effective superfluid density component  $n^{\rs,\rme}_{\alpha\alpha}$ in units of $n^{\rs\para}_{\alpha\alpha}$ as a function of $n^{\rs\perp}_{\alpha\alpha}/n^{\rs\para}_{\alpha\alpha}$ for the spaghetti phase in the absence of entrainment. As in Fig. 1,  the full line shows the result of the effective medium theory, the dashed line  the Voigt approximation, and the dotted line  the Reuss approximation.}
\label{Spaghetti}
\end{figure}

In the case of the spaghetti phase, one expects flow to be impeded in the perpendicular direction ( $n_{\alpha\alpha}^{\rs\perp} <n_{\alpha\alpha}^{\rs\para}$), and it is therefore convenient to write
Eq.\ (\ref{1component}) in the form
\be
\frac{n_{\alpha\alpha}^{\rs\rme}}{n_{\alpha\alpha}^{\rs\para}}=  \left[  \frac12  \frac{n_{\alpha\alpha}^{\rs\perp}}{n_{\alpha\alpha}^{\rs\para}} +\frac{1}{16}\left( \frac{n_{\alpha\alpha}^{\rs\perp}}{n_{\alpha\alpha}^{\rs\para}}\right)^2 \right]^{1/2}  + \frac14 \frac{n_{\alpha\alpha}^{\rs\perp}}{n_{\alpha\alpha}^{\rs\para}}  ,
\label{}
\ee
which we plot in Fig.\ \ref{Spaghetti}.  For small $n_{\alpha\alpha}^{\rs\perp}$, $n_{\alpha\alpha}^{\rs\rme}\simeq  \sqrt{  n_{\alpha\alpha}^{\rs\perp}n_{\alpha\alpha}^{\rs\para}/2}$, which vanishes for $ n_{\alpha\alpha}^{\rs\perp} \to 0$.

\subsection{Effects of entrainment}

On writing out the matrix products in Eq.\ (\ref{f}) explicitly, one finds
\begin{widetext}
\be
f^\lambda_{\rn\rn}=\frac{   -n_{\rn\rn}^{\rs,\rme} [n^{\rs\lambda}_{\rn\rn}n^{\rs\lambda}_{\rp\rp}-(n^{\rs\lambda}_{\rn\rp})^2] - 2 n^{\rs,\rme}_{\rn\rn} (n^{\rs,\rme}_{\rp\rp} n^{\rs \lambda}_{\rn\rn}+n^{\rs,\rme}_{\rn\rp} n^{\rs \lambda}_{\rn\rp} ) +
 (n^{\rs,\rme}_{\rn\rn})^2 n^{\rs \lambda}_{\rp\rp}+ 3 (n^{\rs,\rme}_{\rn\rp})^2  n^{\rs \lambda}_{\rn\rn}  + 2 n^{\rs,\rme}_{\rn\rn} [n^{\rs,\rme}_{\rn\rn} n^{\rs,\rme}_{\rp\rp} - (n^{\rs,\rme}_{\rn\rp})^2 ] }{(n^{\rs\lambda}_{\rn\rn} + 2 n^{\rs,\rme}_{\rn\rn}) (n^{\rs\lambda}_{\rp\rp} + 2 n^{\rs,\rme}_{\rp\rp}) -(n^{\rs\lambda}_{\rn\rp} +2n^{\rs,\rme}_{\rn\rp})^2 },
 \ee
 \be
f^\lambda_{\rp\rp}=\frac{   -n_{\rp\rp}^{\rs,\rme} [n^{\rs\lambda}_{\rn\rn}n^{\rs\lambda}_{\rp\rp}-(n^{\rs\lambda}_{\rn\rp})^2] - 2 n^{\rs,\rme}_{\rp\rp} (n^{\rs,\rme}_{\rn\rp} n^{\rs \lambda}_{\rn\rp} + n^{\rs,\rme}_{\rn\rn} n^{\rs \lambda}_{\rp\rp})+
 (n^{\rs,\rme}_{\rp\rp})^2 n^{\rs \lambda}_{\rn\rn} + 3 (n^{\rs,\rme}_{\rn\rp})^2  n^{\rs \lambda}_{\rp\rp}  + 2 n^{\rs,\rme}_{\rp\rp} [n^{\rs,\rme}_{\rn\rn} n^{\rs,\rme}_{\rp\rp} - (n^{\rs,\rme}_{\rn\rp})^2 ] }{(n^{\rs\lambda}_{\rn\rn} + 2 n^{\rs,\rme}_{\rn\rn}) (n^{\rs\lambda}_{\rp\rp} + 2 n^{\rs,\rme}_{\rp\rp}) -(n^{\rs\lambda}_{\rn\rp} +2n^{\rs,\rme}_{\rn\rp})^2 },
 \ee
 and
 \be
f^\lambda_{\rn\rp}=\frac{ -n_{\rn\rp}^{\rs,\rme} [n^{\rs\lambda}_{\rn\rn}n^{\rs\lambda}_{\rp\rp}-(n^{\rs\lambda}_{\rn\rp})^2] - 3 n^{\rs,\rme}_{\rn\rn}  n^{\rs,\rme}_{\rp\rp} n^{\rs \lambda}_{\rn\rp} + n^{\rs,\rme}_{\rn\rp} (n^{\rs,\rme}_{\rn\rn}  n^{\rs \lambda}_{\rp\rp} +  n^{\rs,\rme}_{\rp\rp} n^{\rs\lambda}_{\rn\rn} ) + (n^{\rs,\rme}_{\rn\rp})^2 n^{\rs \lambda}_{\rn\rp} +
 2 n^{\rs,\rme}_{\rn\rp} [n^{\rs,\rme}_{\rn\rn}   n^{\rs,\rme}_{\rp\rp} - (n^{\rs,\rme}_{\rn\rp})^2] }
{(n^{\rs\lambda}_{\rn\rn} + 2 n^{\rs,\rme}_{\rn\rn}) (n^{\rs\lambda}_{\rp\rp} + 2 n^{\rs,\rme}_{\rp\rp}) -(n^{\rs\lambda}_{\rn\rp} +2n^{\rs,\rme}_{\rn\rp})^2 }.
 \ee
 \end{widetext}
For the uniaxial case, the effective superfluid densities $n^\rme_{\alpha\beta}$ are the solutions of the three simultaneous equations
\be
f^\parallel_{\alpha\beta} +2f^\perp_{\alpha\beta}=0.
\ee
To leading order in $n^{\rs\lambda}_{\rn\rp}$,  the nn and pp components of the effective superfluid density tensor  are given by Eq.\ (\ref{1component}) with the densities $n^{\rs\lambda}$ and $n^{\rs,\rme}$ put equal to the nn and pp components:
\be
n_{\rn\rn}^{\rs,\rme}  \simeq \frac{n_{\rn\rn}^{\rs \perp}+\sqrt{(n_{\rn\rn}^{\rs \perp})^2+8n_{\rn\rn}^{\rs \para} n_{\rn\rn}^{\rs \perp}}}{4},
\label{1componentn}
\ee
\be
n_{\rp\rp}^{\rs,\rme}\simeq   \frac{n_{\rp\rp}^{\rs,\perp}+\sqrt{(n_{\rp\rp}^{\rs\perp})^2+8n_{\rp\rp}^{\rs \para} n_{\rp\rp}^{\rs \perp}}}{4},
\label{1componentp}
\ee
and
\be
n^{\rs,\rme}_{\rn\rp}  \simeq 3  n^{\rs,\rme}_{\rn\rn}n^{\rs,\rme}_{\rp\rp} \frac{ \Delta^\perp n^{\rs\para}_{\rn\rp}+2\Delta^\para n^{\rs\perp}_{\rn\rp}}{   \Delta^\perp \Gamma^\para+2\Delta^\para \Gamma^\perp}   .
\label{nnpapprox}
\ee
Here
\bea
\Delta^\lambda=(n^{\rs\lambda}_{\rn\rn}+2 n^{\rs,\rme}_{\rn\rn})(n^{\rs\lambda}_{\rp\rp}+2 n^{\rs.\rme}_{\rp\rp}),\,\,\,{\rm and}\;\;\; \label{Delta}\\
\Gamma^\lambda =  -n^{\rs\lambda}_{\rn\rn} n^{\rs\lambda}_{\rp\rp} +  n^{\rs\lambda}_{\rn\rn}n^{\rs,\rme}_{\rp\rp}+n^{\rs,\rme}_{\rn\rn}n^{\rs\lambda}_{\rp\rp}+2n^{\rs,\rme}_{\rn\rn}n^{\rs,\rme}_{\rp\rp}.\label{Gamma}
\eea
In Eqs.\ (\ref{nnpapprox})-(\ref{Gamma}), the $n^{\rs,\rme}_{\alpha\alpha}$ should be taken to be the values (\ref{1componentn}) and (\ref{1componentp}) in the absence of entrainment.  For an isotropic inclusion, $n^{\rs,\rme}_{\rn\rp}$ is equal to $n^{\rs\para}_{\rn\rp}=n^{\rs\perp}_{\rn\rp}$, as one would expect physically.

Once the effective superfluid density tensor has been determined, the effective normal densities of neutrons and protons may be found from the condition for Galilean invariance, Eq.\ (\ref{Galileo}).

It is interesting to consider the case of no normal currents ($\vec u=0$) and no bulk flow of protons.  The neutron current density is then given by
\be
\langle j_\rn^i\rangle= \left(n_{\rn\rn}^{\rs, \rme} -    \frac{(  n_{\rn\rp}^{\rs, \rme}  )^2}{n_{\rp\rp}^{\rs, \rme}}        \right) \frac{\langle\nabla^i\phi_\rn\rangle}{m} .
\label{currentalphaav}
\ee
The second term in parentheses reflects the fact that backflow of neutrons around an exclusion results in backflow of protons.  To achieve no average proton current, a nonzero value of   $\langle\nabla^i\phi_\rp\rangle$ is required.
\section{Discussion}

It is not possible to make detailed predictions for the effective superfluid density tensor for disordered pasta phases because there are to date no detailed calculations of the superfluid density tensor for perfectly ordered phases.  Even basic issues such as the thermodynamic stability of the pasta phases need to be addressed with the use of state of the art nuclear physics methods.  However, a number of remarks can be made.  Because the effects of entrainment are not expected to be large, we shall consider the case when it is absent.  Our calculations predict a qualitative difference between the lasanga and spaghetti phases, since even if the superfluid density for flow perpendicular to lasagna sheets is zero, the effective superfluid density drops to only one half of the value for flow in directions in the plane of the lasanga sheets. For spaghetti, when the superfluid density is zero for flow perpendicular to the strands vanishes, the effective superfluid density is zero.  This difference is due to the fact that, while the lasanga phase has two ``easy'' axes (those lying in the plane of the sheets) and one ``hard'' axis, the spaghetti phase has one ``easy'' axis (along the strands) and two ``hard'' axes. We therefore conclude that for the lasagna phase, the effective superfluid density for the disordered state will be greater than one half of the superfluid density tensor for flow in the plane of the sheets.  However, the effective superfluid density for the disordered spaghetti state depends more sensitively on the component of the superfluid density tensor for flow perpendicular to the strands.

Another prediction of our calculations is that for the spaghetti phase, the situation for protons will be different from that for neutrons.  For neutrons, superfluid flow perpendicular to the strands is relatively easy because of the neutron fluid between strands, but for protons in the perfectly ordered spaghetti phase, flow between strands occurs by tunneling and is consequently small. The effective proton superfluid density of the disordered spaghetti phase is therefore sensitive to imperfections such as bridges between adjacent strands.
The effective medium approach has been very successful in accounting for the elastic and electrical properties of terrestrial materials, as may be seen from, e.g., Ref.\ \cite{Landauer} for binary metallic mixtures, but it would be useful to explore how well it works for very anisotropic phases.

A basic assumption of the effective medium approach is that the system may be regarded as a collection of ``domains'' which are oriented  randomly, a situation explored in the case of elastic constants in the simulations of Ref.\ \cite{Caplan}.   An important question is whether or not this assumption is realistic, since it could well be that the orientation of the pasta varies continuously in space, rather than having sharp boundaries between regions with different orientations.  It would be valuable to make simulations to investigate the disorder of pasta structure that results when matter is cooled below the critical temperature for formation of pasta structures.

In the calculations above we have largely neglected the effects of magnetic fields.  In a charged superfluid, the length scale for variations of the magnetic field is of order the London length, $\lambda=(4 \pi n_\rp e^2/mc^2)^{-1/2}$, which is of order $110 $ fm  for a proton fraction of 5\% and a density of half nuclear saturation density.  A spatially independent vector potential $\vec A$ has no physical effect, and in the formalism can be taken into account by shifting the proton phase by an amount $e\vec A\cdot \vec r/c$.  Only the spatially dependent part of the vector potential is physically relevant, and in order to be able to neglect this, the size of the inclusion, which corresponds to the spatial scale of the disorder of the orientation of the pasta elements,  must be less than the London length.  This requires that the pasta be rather highly disordered, since, e.g., for the lasanga phase the lattice spacing is estimated to be about 44 fm \cite{ZhangCJP}.

Superconducting vortices in ordered pasta phases have been discussed in Ref.\ \cite{ZhangCJP}, and the vortex energy depends on its direction with respect to the principle axes.  In disordered phases, the properties of vortices will become independent of direction if the length scale of the disorder becomes small compared  with the superconducting penetration depth.

\section{Conclusion}
We have developed an effective medium approach to calculating the properties of disordered pasta phases that allows for the effects of entrainment.
For the lasagna phase, the calculations predict that, with the neglect of entrainment, the diagonal (nn and pp) components of the effective superfluid density are greater than one half of the corresponding value of the superfluid density tensor for flow in directions lying in the plane of the sheets.
Because flow of protons perpendicular to the direction of the strands is suppressed in the spaghetti phase, the pp component of the effective superfluid density tensor can be very much less than the corresponding value of the superfluid density tensor for flow in the direction of the strands.   However, since neutrons can move relatively easily between strands, the reduction of the nn component of the effective superfluid density tensor is expected to be much less.

In our calculations we have not taken into account  possible effects of the interface, e.g., the existence of weak links between neighboring regions with different orientations of the pasta structures.  Such effects are important for laboratory superconductors because of the existence of grain boundaries \cite{ClemKogan} but are expected to be less important in neutron star crusts where on length scales comparable to the internucleon spacing the structure is that of a liquid rather than a solid.   One situation in which interface effects could be important for protons is if there is no path for percolation within the proton-rich phase.  However, for neutrons, which are present everywhere, such effects are unlikely to be important.

To improve estimates of the effective superfluid density of the disordered pasta phases, it is necessary to have better values for the superfluid density tensor for the ordered phases.   One possible approach would be to extend the calculations of Ref.\ \cite{WatanabeCJP} for fermions in a one-dimensional sinusoidal potential to potentials that better reflect the structure of the pasta phases.

\section{Acknowledgements}
We are grateful to Brian M\o ller Andersen, Ruslan Prozorov, and David Stroud for helpful correspondence.

\appendix
\section{Simple derivation of the basic result}

Consider an inclusion consisting of the pasta phases with superfluid density tensor $n_{\alpha\beta}^{\rs, ij}$ immersed in an isotropic medium with superfluid density tensor $n_{\alpha\beta}^{\rs, \rme}$.   For simplicity, we shall take the inclusion to be spherical, but the final results we derive also apply for an ellipsoidal inclusion when the principle axes of the ellipsoid coincide with the principle axes of the superfluid density tensor; this may be demonstrated by working in terms of ellipsoidal coordinates rather than spherical polar coordinates, just as in the analogous problem in electrostatics \cite{LandLElectrodynamics}.
In the uniform medium without the inclusion, the current density in the presence of uniform gradients of the phases given by $\nabla^i \phi_\alpha=k^i_\alpha$ may be written as $j^i_\alpha=n_{\alpha\beta}^{\rs, \rme} k^j_\beta/m$.  The inclusion induces backflow around it and changes the current density within the inclusion.   Because, by symmetry, the current density integrated over space for $\vec k$ lying along one of the principle axes also lies in the direction of that axis,
 it is convenient to work in terms of the components of $\vec k$ along the principle axes of the inclusion, which we denote by the label $\lambda =1, 2, 3$ and we denote the components of the superfluid density tensor by $n_{\alpha\beta}^{\rs \lambda}$.

We consider a spherical inclusion  of a pasta phase with superfluid density tensor $n^{\rs,ij}_{\alpha\beta}$ embedded in a homogeneous medium with effective superfluid density tensor
$n^{\rs,\rme}_{\alpha\beta}$.  We shall denote the component of $\vec k$ along the $\lambda$ axis by $k^\lambda$, and since the problem we are considering is linear, we may superimpose solutions for $\vec k$ along the three principle directions of the inclusion.  In a homogeneous medium, $\phi={\vec k}\cdot {\vec r}$, and the wave vector of the flow is $\vec\nabla \phi =\vec k$ is uniform.   which we shall take to be in the $z$ direction.  The current density is given by Eq.\ (\ref{currentalpha})
and therefore in regions where the superfluid density tensor is isotropic, conservation of particle number demands that
\be
\nabla^i  j^i_\alpha=0,
\label{currentcons}
\ee
and thus $\nabla^2\phi_\alpha=0$.
We shall seek a solution of the form
\be
\phi_\alpha= \left\{\begin{array}{l}        (k^\lambda_\alpha +A^\lambda_\alpha) z\,\,\,\,\,\,\;\;\; {\rm for}\,\,\, r\leq a, \\     k^\lambda_\alpha z+{C^\lambda_\alpha z}/{r^3}\,\,\, {\rm for}\,\,\, r\geq a,  \end{array}                    \right.
\label{solution}
\ee
where the origin of the coordinate system is at the center of the inclusion, $r$ is the radial coordinate, and $z$ is the coordinate in the direction of the principle axis $\lambda$.
From the condition that $\phi_\alpha$ be continuous at the surface of the inclusion ($r=a$), it follows that
\be
A^\lambda_\alpha=\frac{C^\lambda_\alpha}{a^3}.
\ee
Within the inclusion, the current density is constant in space, and thus the solution (\ref{solution}) satisfies the condition for particle conservation (\ref{currentcons}) everywhere except at the surface of the inclusion.  To ensure conservation of particle number there, the component of the current density, Eq.\ (\ref{currentalpha}), normal to the surface must be continuous, which gives
\be
n_{\alpha\beta}^{\rs \lambda}(k^\lambda_\beta+A^\lambda_\beta)=n_{\alpha \beta}^{\rs,\rme } (k^\lambda_\beta-2A^\lambda_\beta   k^\lambda_\beta),
\ee
where we use the summation convention for subscripts but not for the superscript $\lambda$.  Thus in a compact matrix notation,
\be
A^\lambda=   \left( 2 n^{\rs,\rme}+n^{\rs \lambda}       \right)^{-1} \left(    n^{\rs, \rme} -n^{\rs\lambda}   \right)  k^\lambda,
\ee
where $A^\lambda$ and $k^\lambda$ are vectors and $  n^{\rs \lambda}$ and $ n^{\rs,\rme} $ are matrices in the two-dimensional species space.
The current density integrated over the whole of space is given in this matrix notation by
\be
 \int_Vd^3r\,  j^\lambda({\vec r})=    Vn^{\rs,\rme} \frac{k^\lambda}{m}  +  \frac{4\pi a^3}{3} \left[  n^{\rs \lambda} \frac{A^\lambda}{m} -(n^{\rs,\rme} -n^{\rs \lambda})\frac{k^\lambda}{m}\right],
\ee
where $V$ is the volume of the system.  The change in the integrated current density due to the presence of the inclusion is therefore
\be
\Delta \int_Vd^3r\,  j^\lambda({\vec r})    = \frac{4\pi a^3}{3} \left[  n^{\rs\lambda} \frac{A^\lambda}{m} -(n^{\rs,\rme} -n^{\rs \lambda})\frac{k^\lambda}{m}\right]\\
\ee
\be
= -6V_\rmi n^{\rs,\rme}  (2n^{\rs, \rme}+ n^{\rs \lambda})^{-1}  (n^{\rs,\rme} -n^{\rs \lambda})\frac{k^\lambda}{m},
\ee
where
\be
V_\rmi =\frac{4\pi a^3}{3}
\ee
is the volume of the inclusion.  While locally the current density has components in directions other than $\lambda$, the integrals over space of these components vanish because of the dipolar character of the flow outside the inclusion.

For $\vec k$ in an arbitrary direction with respect to the principle axes of the inclusion, the change in the integrated current density is given by
\be
\Delta \int_Vd^3r\,  {\vec j}({\vec r})= -6V_\rmi \sum_{\lambda=1,2,3}f^\lambda\frac{\vec k\cdot {\vec\epsilon}^\lambda}{m} {\vec\epsilon}^\lambda,
\ee
 where
 \be
 f^\lambda= n^{\rs,\rme}  (2n^{\rs, \rme}+ n^{\rs \lambda})^{-1}  (n^{\rs,\rme} -n^{\rs \lambda})
 \label{f}
 \ee
and  $ {\vec\epsilon}^\lambda$ is a unit vector in the $ \lambda$ direction.  On averaging over possible orientations of the principle axes of the inclusion, the current density integrated over space is in the direction of $\vec k$
 and
 \be
\langle \Delta \int_Vd^3r\,  {\vec j}({\vec r})\cdot\vec k\rangle= \frac{6V_\rmi}{m} \sum_{\lambda=1,2,3}f^\lambda \langle(\vec k\cdot {\vec\epsilon}^\lambda)^2\rangle.
 \ee
For a random orientation of the grains, $\langle(\vec k\cdot {\vec\epsilon}^\lambda)^2\rangle$ is equal to $k^2/3$ and the integrated current density vanishes when
\be
\sum_{\lambda=1,2,3}f^\lambda=0,
\label{selfconsist}
\ee
which is the condition determining $n^{\rs,\rme}$ in the self-consistent effective medium approach.  For the two-component case, $f$ is a $2\times 2$ symmetric matrix, and therefore this equation is equivalent to three scalar equations.

\end{document}